\documentclass[prl,preprint,aps,superscriptaddress,showpacs,floatfix]{revtex4}
\usepackage{amsmath}
\usepackage{graphicx}

\begin{document}

\title{Layered superconductors as negative-refractive-index metamaterials}
\author{A.L. Rakhmanov}
\affiliation{Advanced Science Institute, The Institute of Physical
and Chemical Research (RIKEN), Saitama, 351-0198, Japan}
\affiliation{Institute for Theoretical and Applied Electrodynamics
Russian Acad. Sci., 125412 Moscow, Russia}

\author{V.A. Yampol'skii}
\affiliation{Advanced Science Institute, The Institute of Physical
and Chemical Research (RIKEN), Saitama, 351-0198, Japan}
\affiliation{A.Ya.~Usikov Inst. for Radiophysics and Electronics,
Ukrainian Acad. Sci., 61085
Kharkov, Ukraine}

\author{J.A. Fan}
\affiliation{School of Engineering and Applied Sciences, Harvard University, Cambridge, Massachusetts 02138, USA}

\author{Federico Capasso}
\affiliation{School of Engineering and Applied Sciences, Harvard University, Cambridge, Massachusetts 02138, USA}

\author{Franco Nori}
\affiliation{Advanced Science Institute, The Institute of Physical
and Chemical Research (RIKEN), Saitama, 351-0198, Japan}
\affiliation{Department of Physics, University of
Michigan, Ann Arbor, MI 48109-1040, USA}

\begin{abstract}
We analyze the use of layered superconductors as anisotropic metamaterials.
Layered superconductors can have a negative refraction index in a wide frequency
range for arbitrary incident angles. Indeed, low-$T_c$ ($s$-wave) superconductors allow to
produce artificial heterostructures with \textit{low losses} for $T\ll T_c$. However,
the real part of their in-plane effective permittivity is very large. Moreover, even at low temperatures,
layered high-$T_c$ superconductors have a large in-plane normal conductivity, producing large
losses (due to $d$-wave symmetry). Therefore, it is difficult to enhance the evanescent
modes in either low-$T_c$ or high-$T_c$ superconductors.
\end{abstract}

\maketitle

Metamaterials are attracting considerable attention because of their unusual interaction with electromagnetic waves (see e.g.,~\cite{ves}). In particular, metamaterials supporting negative refractive index have the potential for subwavelength resolution~\cite{pen1} and aberration-free imaging. The first proposed negative index metamaterials used subwavelength electric and magnetic structures to achieve simultaneously negative permittivity $\varepsilon$ and permeability $\mu$ (see, e.g.,~\cite{shalaevreview}).  However, these ``double negative" structures require intricate design and demanding fabrication techniques, are not ``very subwavelength", and suffer from spatial dispersion effects.  Moreover, the implicit overlapping electric and magnetic resonances (see, e.g.,~\cite{over}) often leads to resonant losses that, together with material losses, lead to significant degradation in metamaterial functionality.  This manifestation of loss can be quantified by examining the figure of merit (FOM) in such materials, which is defined as $|n'|/n''$ where $n'$ and $n''$ are the real and imaginary parts of the refractive index $n$, respectively.  The FOMs of negative index materials in the visible and near-IR have experimentally ranged from 0.1 up to 3.5~\cite{shalaevreview,threeDmaterial}.

Another promising route to creating negative index metamaterials is to use strongly anisotropic materials, in particular, uniaxial anisotropic materials with different signs of the permittivity tensor components along, $\varepsilon_\|$, and transverse, $\varepsilon_\bot$, to the surface (see, e.g.,~\cite{anisa1,anisa2}).  These materials have been theoretically~\cite{narimanov1} and experimentally~\cite{hyperlens} demonstrated to support sub-wavelength imaging, and they have also been proposed as a model system for scattering-free plasmonic optics~\cite{scatterfree} and subwavelength-scale waveguiding~\cite{funnel}.  These materials are particularly attractive because: they are relatively straightforward to fabricate, compared to double negative metamaterials; they do not require negative permeability; and do not suffer from magnetic resonance losses.  The FOMs for such materials have been calculated to be significantly greater than those measured in double negative materials~\cite{hoffman, anisa1}.

Experimental schemes for creating strongly anisotropic uniaxial materials have typically involved the fabrication of subwavelength stacks of materials whose layers comprise alternating signs of permittivity.  For example, alternating stacks of Ag and Al$_2$O$_3$~\cite{hyperlens} and of doped and undoped semiconductors~\cite{hoffman} have been demonstrated to support strong anisotropy in the visible and infrared frequency ranges respectively. However, spatial dispersion can strongly modify the optical response of the system relative to the ideal effective medium limit response~\cite{spatialdispersion}; strong local field variations exist due to the structure and length scales of plasmonic modes supported by negative-permittivity films, even in the limit of $l \ll a_{0}$, where $l$ is the length scale of the thin films in the material and $a_{0}$ is the free-space electromagnetic wavelength. This imposes limitations to subwavelength imaging and waveguiding in such materials.  Spatial dispersion may be reduced by making composite structures with thinner layers. However, there exist practical material deposition limitations to thin-film stacks involving film roughness and continuity. In addition, damping due to electron scattering at the thin film interface becomes significant starting at length scales of $a_{0} v_{\rm F}/c \sim a_{0}/100$ where $v_{\rm F}$ is the Fermi velocity in the material~\cite{funnel2}, limiting the minimum film thickness.  It is clear that composite structures are limited in practice as ``ideal" strongly anisotropic materials.

We analyze here the idea of using superconductors as metamaterials (see, e.g.,~\cite{supra,supra1,cap}). In particular, we consider layered cuprate superconductors~\cite{cap} and artificial superconducting-insulator systems~\cite{pim} as candidates for strongly anisotropic metamaterials.  Unlike the composite structures discussed earlier, layered superconductors are not limited in performance by the spatial dispersion effects discussed in~\cite{spatialdispersion}.  We will analyze these materials in the specific context of subwavelength resolution, which can be achieved by the amplification of evanescent waves~\cite{pen1}. This amplification is high when $n$ is close to unity and its imaginary part is small~\cite{pen1,ss}. For the incident $p$-polarized waves considered here, subwavelength resolution requires $\textrm{Im} (\varepsilon)\ll\exp(-2k_\bot L)$, where $k_\bot$ is the wavevector component across the surface, and $L$ is the plane lens-thickness~\cite{ss}. For evanescent modes with $k_\bot=2\omega/c=2k_0$ and $L/a_0=0.1$, we have $\textrm{Im} (\varepsilon)\ll 0.081$.

We show that in the case of natural high-$T_c$ cuprates the losses are high at any reasonable frequency. In the case
of artificial layered structures prepared from low-$T_c$ superconductors, the losses can be
reduced significantly at low temperatures, $T\ll T_c$, where $T_c$ is the critical
temperature. The frequency range for such a metamaterial
is $\hbar\omega<2\Delta$, where $\Delta$ is the superconducting gap, which
corresponds to a maximum frequency in the THz range for low-$T_c$ superconductors.
We prove that the in-plane permittivity for low-$T_c$ multi-layers is large, preventing
the effective enhancement of evanescent waves. This is problematic because subwavelength
resolution~\cite{pen1} requires the amplification of evanescent waves.
Note that Refs.~\onlinecite{supra1} only focus on the zero-frequency DC case.

\textit{Effective permittivity}.--- We study a medium consisting of a periodic stack of superconducting layers
of thickness $s$ and insulating layers of thickness $d$ with Josephson coupling between successive superconducting planes.
The number of layers is large, $L/(s+d)=N\gg 1$ and $s$ is smaller than: the in-plane magnetic field penetration
depth $\lambda_{\|}$, transverse skin depth $\delta_{\bot}(\omega)$, and wavelength $a(\omega)\sim 2\pi c/\omega\sqrt{|\varepsilon(\omega)|}$.
We calculate the effective permittivity, $\widehat{\varepsilon}=(\varepsilon_{\|},\varepsilon_{\bot})$, of the layered system in the case of
$p$-wave refraction.

Layered superconductors with Josephson couplings can be described by the Lawrence-Doniach model,
where the averaged current components can be expressed as~\cite{ka}
\begin{equation}\label{J}
J_{\bot} = J_c\sin{\varphi_n}+ \frac{\sigma_{\bot}\Phi_0\dot{\varphi}_n}{2\pi c(s+d)}\,,\,\,
  J_{\|} = \frac{c\Phi_0p_n}{8\pi^2\lambda^2_{\|}}+\sigma_{\|}E_{\|},
\end{equation}
where $\varphi_n$ is the gauge-invariant phase difference between the $(n+1)$th and
$n$th superconducting layers, $p_n$ is the in-plane superconducting
momentum, $J_c=c\Phi_0/(8\pi^2d\lambda^2_{\bot})$ is the transverse supercurrent density,
$\Phi_0$ is the magnetic flux quantum, and $\lambda_{\bot}$ is the transverse
magnetic field penetration depths. Also $\sigma_{\bot}$ and $\sigma_{\|}$ are the averaged transverse and
in-plane quasiparticle conductivities. The transverse $E_{\bot}$ and
in-plane $E_{\|}$ components of the electric field are related to the
gauge-invariant phase difference and superconducting momentum by~\cite{ka,thz}
\begin{equation}\label{E}
  \left(1-\alpha\nabla^2_n\right)E_{\bot} = \frac{\Phi_0}{2\pi c(s+d)}\dot{\varphi}_n,\,\,\,
  E_{\|} =  \frac{\Phi_0}{2\pi c}\dot{p}_n,
\end{equation}
where $\nabla^2_n f(n)\!=\!f(n+1)\!+\!f(n-1)\!-\!2f(n)$,
$\alpha=\varepsilon R_D^2/(sd)$ is the capacitive coupling between layers, and $R_D$ is
the Debye length. We linearize the first of Eqs.~\eqref{J} and consider a linear electromagnetic wave $\,$
$\!\!E_{\|,\bot}(x,n,t)\!=\!\sum_q\!\!\int\!\!\frac{dk\,d\omega}{(2\pi)^2}E_{\|,\bot}(k,q,\omega)\exp({-i\omega t+ikx+iqn})$,
where $q=\pi l/(N+1)$, $l=0,\pm 1, \pm 2$, and the $x$-axis is in the plane of the layers.
Using Eqs.~\eqref{J} and \eqref{E}, we obtain:
\begin{equation}\label{JE}
\frac{J_{\bot}}{E_{\bot}}\! =\! \left(\!1\!+\!\alpha \widetilde{q^2}\right)\!\!\left[\!\sigma_{\bot}\!-\!\frac{\varepsilon\omega_p^2(s\!+\!d)}{4\pi i d\omega}\!\right], \,\,
\frac{J_{\|}}{E_{\|}}\! =\! \sigma_{\|}\!-\!\frac{\varepsilon\gamma^2\omega_p^2}{4\pi i\omega},
\end{equation}
where $\omega_p=c/(\lambda_{\bot}\sqrt{\varepsilon})$ is the Josephson plasma frequency,
$\varepsilon$ is the interlayer permittivity, $\gamma=\lambda_{\bot}/\lambda_{\|}$,
and $\widetilde{q^2}=2(1-\cos{q})$. Averaged over the sample volume, the Maxwell equation has the form
$c\nabla\times\mathbf{H}=4\pi\mathbf{J}+\partial \mathbf{D}/\partial t$,
where $D_{\|}=\varepsilon^0_{\|}E_{\|}$ and $D_{\bot}=\varepsilon^0_{\bot} E_{\bot}$. In the
effective medium approximation, the components of the permittivity tensor can be expressed as~\cite{LL}
$\varepsilon^0_{\|}=(d\varepsilon+s)/(s+d)$, and $\varepsilon^0_{\bot}=\varepsilon(s+d)/(s\varepsilon+d)$,
where we assume that $\varepsilon_{\textrm{superconductor}}=1$.
Fourier transforming the above Maxwell equation, we derive
$c\left[\nabla\times\mathbf{H}\right]_{\bot}(k,q,\omega)= -\varepsilon_{\bot}E_{\bot}$ and $c\left[\nabla\times\mathbf{H}\right]_{\|}(k,q,\omega)=-\varepsilon_{\|}E_{\|}$,
where $\varepsilon_{\|}=\varepsilon^0_{\|}-(4\pi/i\omega)(J_{\|}/E_{\|})$ and
$\varepsilon_{\bot}=\varepsilon^0_{\bot}-(4\pi/i\omega)(J_{\bot}/E_{\bot})$.
Therefore, we finally obtain
\begin{eqnarray}\label{perm}
\nonumber
\varepsilon_{\bot} &=& \varepsilon^0_{\bot}-\frac{4\pi\left(1+\alpha \widetilde{q^2}\right)\sigma_{\bot}}{i\omega}-\varepsilon\left(1+\alpha \widetilde{q^2}\right)\frac{\omega_p^2(s+d)}{\omega^2d}\,, \\
\varepsilon_{\|} &=& \varepsilon^0_{\|}-\frac{4\pi\sigma_{\|}}{i\omega}-\varepsilon\gamma^2\frac{\omega_p^2}{\omega^2}\,.
\end{eqnarray}
Thus, $\varepsilon_\|<0$ and $\varepsilon_\bot>0$ in the frequency range
\begin{equation}\label{meta}
\sqrt{\left(1+\alpha \widetilde{q^2}\right)\frac{s\varepsilon+d}{d}}<\frac{\omega}{\omega_p}<\gamma\sqrt{\frac{\varepsilon(s+d)}{d\varepsilon+s}}.
\end{equation}
If the incident angle is close to normal and anisotropy is large, $\gamma\gg 1$, we can find an estimate $\textrm{FOM}\approx 2\left|\textrm{Re}(\varepsilon_\|)/\textrm{Im}(\varepsilon_\|)\right|
\approx\varepsilon\gamma^2\omega^3/2\pi\sigma_\|\omega_p^2$. Electromagnetic waves propagate in the layered superconductors if $\omega>\omega_p$.
Thus, the results obtained are valid if $\omega_p<\omega<\omega_c=2\Delta/\hbar$. Below
we analyze separately the different cases of a typical high-$T_c$ layered superconductor, Bi$_2$Sr$_2$CaCu$_2$O$_{8+\delta}$ (Bi2212),
and also an artificial low-$T_c$ layered structure made from Nb.

\textit{Layered high-$T_c$ superconductors}.--- In the case of Bi2212, it is known that
$s\ll d=1$--2~nm, $\varepsilon=12$, $\alpha\approx 0.1$, and at low temperatures ($T\ll T_c=90$~K)
$\omega_p\approx 10^{12}$~s$^{-1}$, $\gamma=500$, $\sigma_{\|}\approx 4\cdot 10^4$~
$\Omega^{-1}$cm$^{-1}$, and $\sigma_{\bot}\approx 2\cdot 10^{-3}$~$\Omega^{-1}$cm$^{-1}$~\cite{ka,la}.
In this case, Eqs.~\eqref{perm} can be rewritten as $\varepsilon_{\bot}\! \approx\! \varepsilon\!\left(1\!-\!\omega^2/\omega_p^2\right)\!+\!4\pi i\sigma_{\bot}/\omega$,
$\varepsilon_{\|}\! \approx\! \varepsilon\!\left(1\!-\!\gamma^2\omega^2/\omega_p^2\right)\!+\!4\pi i\sigma_{\|}/\omega$.
The calculated frequency dependence of the permittivity for Bi2212 is shown in Fig.~1. The superconducting gap for Bi2212 is estimated as
$\Delta\approx 2$-$3 k_BT_c$, with $\omega_c\approx 5\times 10^{13}$~s$^{-1}\ll \gamma\omega_p$.
Thus, for any incident angle, Bi2212 has negative $n$ in the frequency range from
about 0.15~THz to 7.5~THz, or in the wavelength domain 40~$\mu$m$\,\lesssim a\lesssim 2$~mm.
However, the use of Bi2212 as metamaterial has a disadvantage since the in-plane quasiparticle conductivity
$\sigma_\|$ is large, even at helium temperatures. As it is seen from the inset in Fig.~1(b), $\sigma_\|\neq 0$ when $T\rightarrow 0$,
which is typical for superconductors having a $d$-type symmetry of the order parameter. In addition,
the usual dimensions of high-quality Bi2212 single crystals are less than 1~mm in the in-plane direction and about 30--100~$\mu$m
in the transverse direction. Thus, it might be difficult to use Bi2212 single crystals as metamaterials, or elements of a superlens.

\begin{figure}[btp]
\begin{center}
  \includegraphics[width=8 cm]{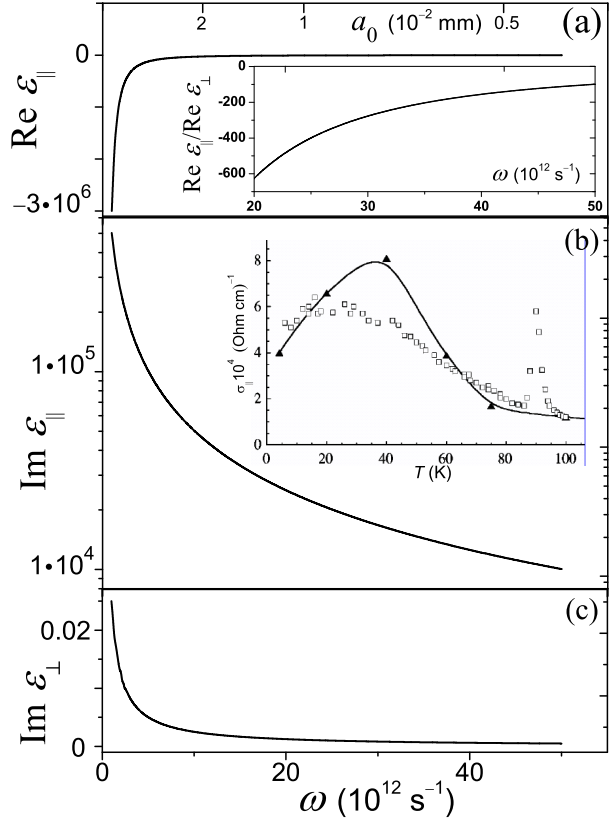}\\
\end{center}
\caption{Dependence of the real and imaginary parts of the permittivity $\hat{\varepsilon}$ in
Bi2212 on frequency $\omega$ (or wavelength $a_0$), calculated from Eqs.~\eqref{perm}: (a) real part of the in-plane permittivity $\varepsilon_\|(\omega)$. Inset: ratio of the real parts of the in-plane and transverse permittivities; (b) imaginary part of the in-plane permittivity. Inset: temperature dependence of the in-plane quasiparticle conductivity $\sigma_{\|}(T)$ in Bi2212; solid triangles are low frequency data from Ref.~\onlinecite{la}, open squares correspond to 14.4~GHz data from Ref.~\onlinecite{Shi}; (c) imaginary part of the transverse permittivity $\varepsilon_\bot(\omega)$.}
\label{Bi}
\end{figure}

\textit{Low-$T_c$ artificial layered structures}.--- The thickness of the insulator in Josephson junctions
is about a few nm. To attain a low-loss regime and reach the bulk critical temperature, the thickness of the
superconducting layers should be larger or about the superconductor coherence length $\xi$. For clean
superconductors, $\xi$ is about tens of nm. Thus, for low-$T_c$ artificial-layered structures, it is reasonable
to analyze the case $d\ll s$. In this limit, $\lambda_{\|}=\lambda\sqrt{(s+d)/s}\approx\lambda$, where $\lambda$ is the bulk magnetic
field penetration depth and $\alpha=\varepsilon R_D^2/(sd)\ll 1$ in any realistic case. It is easy to choose an insulator
with very low conductivity $\sigma_i$ to satisfy the condition $\sigma_i\ll \sigma_sd/s$ at any reasonable temperature,
where $\sigma_s$ is the quasiparticle conductivity of the superconductor. In this case we have $\varepsilon_{\bot}^0=1,\,\,\varepsilon_{\|}^0=1,\,\,\sigma_{\bot}=\sigma_is/d$
and $\sigma_{\|}=\sigma_s$. Equations~\eqref{perm} for the effective permittivity can now be rewritten as
\begin{equation}\label{nb}
\varepsilon_{\bot}\! \approx\! \left(\!\!1\!-\!\varepsilon\frac{s\omega_p^2}{d\omega^2}\!\right)\!+\!\frac{4\pi i\sigma_is}{\omega d},\,\,
\varepsilon_{\|}\! \approx\! \varepsilon\!\left(\!\!1\!-\!\gamma^2\frac{\omega_p^2}{\omega^2}\!\right)\!+\!\frac{4\pi i\sigma_s}{\omega}.
\end{equation}
Therefore,  the refraction index $n$ is negative if
\begin{equation}\label{artM}
\sqrt{\varepsilon s/d}<\omega/\omega_p<\gamma.
\end{equation}
For artificial structures, $\gamma$ can be easily made of the order of, or even much larger than, in natural layered
superconductors. In contrast to $d$-wave high-$T_c$ superconductors, for bulk $s$-wave superconductors,
the quasiparticle conductivity $\sigma_s$ tends to zero for decreasing $T$. Thus, in principle, the imaginary part
of $\varepsilon_\|$ could be made as small as necessary by cooling the system.

\begin{figure}[btp]
\begin{center}
  \includegraphics[width=8 cm]{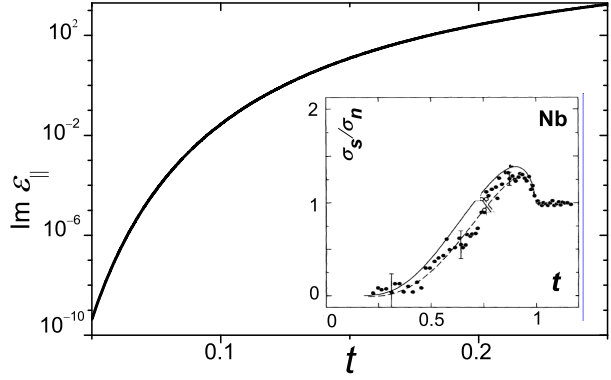}\\
\end{center}
\caption {Calculated, from Eq.~\eqref{mb0}, temperature dependence of the imaginary part of $\varepsilon_\|(t)=\varepsilon_\|(T/T_c)$, for a Nb-based layered structure, with $\omega=0.9\omega_c$, $\varepsilon=10$, $s/d=5$, and $\gamma=500$; here: $\omega_p/\omega_c=0.1$, $\textrm{Re}\;(\varepsilon_\bot)=0.393$, and $\textrm{Re}\;(\varepsilon_\|)\approx-3\cdot 10^{6}$. The inset shows the dependence~\cite{Nb} of $\sigma_s/\sigma_n$ on $t\equiv T/T_c$; points: experimental data for Nb at about 60~GHz; solid line: Mattis-Bardeen theory in the weak-coupling BCS limit; dashed line: strong-coupling Eliashberg prediction~\cite{Nb}.}\label{calc}
\end{figure}

Consider now Nb superconducting layers. For estimates we can take~\cite{Nb}:
$T_c=9.3$~K, $\lambda(T=0)=44$~nm, $\xi=38$~nm, electron mean free path $l_e=20$~nm, and normal state conductivity
$\sigma_n=0.85\times 10^6$~$\Omega^{-1}$cm$^{-1}$. Thus, a reasonable thickness for the superconducting layers can be chosen as
$s=30$--40~nm $\ll a(\omega_c)\gtrsim100$--200~nm. Superconducting properties of Nb are well described in the BCS weak-coupling
approximation~\cite{Nb}. In particular, its conductivity $\sigma_s(\omega,T)$ can be calculated using the Mattis-Bardeen theory~\cite{mb}
(see inset in Fig.~2). At low temperatures, $T\ll T_c$, in the weak-coupling BCS limit, we have $\Delta=1.76\,k_BT_c$.
When $\omega<\omega_c$ and $T\ll T_c$, we can rewrite the Mattis-Bardeen formula for conductivity~\cite{Nb,mb} in the form

\begin{eqnarray}\label{mb0}
\nonumber
\sigma_s/\sigma_n=\omega_c\left[1-\exp\left({-3.52\omega/(\omega_ct)}\right)\right]/\omega\times \\
\int_1^\infty\frac{\left(u^2+1+2u\omega/\omega_c\right)\exp{\left(-\frac{1.76u}{t}\right)}}{\sqrt{\left(u^2-1\right)\left[\left(u+2\omega/\omega_c\right)^2-1\right]}}du,
\end{eqnarray}
where $t=T/T_c$. The results of our calculations are shown in Fig.~\ref{calc}.
These calculations demonstrate that the losses in artificial structures made from low-$T_c$ superconductors \textit{can be extremely low}.
The maximum frequency $\omega_c=3.52\,k_BT_c/\hbar$ for Nb corresponds to approximately 0.7~THz.
From the results presented in Fig.~2, we can estimate that at $\omega\sim\omega_c$ the imaginary part
of $\varepsilon_\|$ is lower than 10$^{-3}$ if $T<1$~K. At higher frequencies, $\omega>\omega_c$,
the conductivity of the superconductor is about the conductivity of the normal metal and it cannot be easily used as a metamaterial with
low losses. Note also that by an appropriate choice of insulator, $s$, and $d$, we can vary
the parameters $\gamma$ and $\omega_p$ in a wide range. If we assume that $\varepsilon\sim 10$, then to fulfill conditions \eqref{artM}
for $\omega_p<\omega_c$ we should prepare highly-anisotropic heterostructures with $\gamma > 10^3$. If the anisotropy is large,
we can find from Eq.~\eqref{nb} that $\textrm{Re}\;(\varepsilon_\|)\approx -c^2/\lambda^2\omega^2$. The absolute value
of $\textrm{Re}\;(\varepsilon_\|)$ is very large, $|\textrm{Re}\;(\varepsilon_\|)|\geq c^2/\lambda^2\omega_c^2\approx 3\times 10^6$. These estimates
suggest that low-$T_c$ superconducting multi-layers might not work as practical metamaterials.

The metamaterial properties of layered superconductors, either natural or artificial,
can be tuned varying the temperature or an in-plane magnetic field,
which strongly affects the transverse critical current density and, consequently, the plasma frequency.
But applying a magnetic field increases dissipation, which is undesirable. Note also that the estimates made above show that the value of FOM
may be very large for the systems considered here, however, this does not mean necessarily that these media can be easily used as practical metamaterials.

\textit{Cuprates in the normal state}.--- There is experimental evidence that cuprate superconductors have strongly anisotropic optical characteristics in the normal state~\cite{norm, norm1}. For example, it was observed that La$_{2-x}$Sr$_x$CuO$_4$ supports negative permittivity along the CuO planes at frequencies up to the mid- and near-IR range~\cite{norm}.  Moreover, these optical properties could be finely tuned by varying the stoichiometry. Such natural materials are thus candidates for practical anisotropic metamaterials. The use of cuprates in the normal state have evident advantages, such as operating above $\omega_c$ and to work at room temperature. However, the normal conductivity of cuprates is of the same order as their quasi-particle conductivity in the superconducting state (see, e.g., the inset in Fig.~1b and Ref.~\onlinecite{la}). The metamaterial properties of cuprates in the normal state require a separate analysis and will be performed elsewhere.

\textit{Conclusions}.--- Here we analyze the properties of anisotropic metamaterials made from layered superconductors. We show that these materials can have a negative refraction index in a wide frequency range for arbitrary incident angles. However, superconducting metamaterials made from natural layered high-$T_c$ cuprates have a large in-plane normal conductivity, even at very low temperatures, due to $d$-wave symmetry of their superconducting order parameter. Therefore, these are very lossy.  Nevertheless, low-$T_c$ $s$-wave superconductors allow to produce metamaterials with \textit{low losses} at low temperatures, $T\ll T_c$. But the real part of their in-plane permittivity is very large, reducing the enhancement of the evanescent modes and potentially limiting the use of superconducting structures as practical metamaterials.

We gratefully acknowledge partial support from the NSA, LPS, ARO, NSF grant No.
EIA-0130383, and JSPS-RFBR 09-02-92114, FC gratefully
acknowledges useful discussions with E.~Narimanov.

\end{document}